\documentclass{aa}  

\usepackage{ifpdf}
\pdfoutput=1
\usepackage{graphicx}
\usepackage{txfonts}
\usepackage[squaren,cdot]{SIunits}
\usepackage{tabularx}
\usepackage[breaklinks=true]{hyperref} 
\usepackage{natbib,twoopt}
\bibpunct{(}{)}{;}{a}{}{,} 
\makeatletter
\newcommandtwoopt{\citeads}[3][][]{\href{http://adsabs.harvard.edu/abs/#3}%
{\def\hyper@linkstart##1##2{}%
\let\hyper@linkend\@empty\citealp[#1][#2]{#3}}}
\newcommandtwoopt{\citepads}[3][][]{\href{http://adsabs.harvard.edu/abs/#3}%
{\def\hyper@linkstart##1##2{}%
\let\hyper@linkend\@empty\citep[#1][#2]{#3}}}
\newcommandtwoopt{\citetads}[3][][]{\href{http://adsabs.harvard.edu/abs/#3}%
{\def\hyper@linkstart##1##2{}%
\let\hyper@linkend\@empty\citet[#1][#2]{#3}}}
\newcommandtwoopt{\citeyearads}[3][][]%
{\href{http://adsabs.harvard.edu/abs/#3}
{\def\hyper@linkstart##1##2{}%
\let\hyper@linkend\@empty\citeyear[#1][#2]{#3}}}
\makeatother

\begin{document}

    \title{The Carlina-type diluted telescope:}
\subtitle{Stellar fringes on Deneb}

\author{
  H. Le Coroller \inst{1,2} \and
  J. Dejonghe    \inst{3} \fnmsep \thanks{First and Second authors contributed equally} \and
  F. Hespeels    \inst{4,1} \and
  L. Arnold      \inst{1} \and
  T. Andersen    \inst{5} \and
  P. Deram       \inst{1} \and
  D. Ricci       \inst{6,1} \and
  P. Berio       \inst{3} \and
  A. Blazit      \inst{3} \and
  J-M. Clausse   \inst{3} \and
  C. Guillaume   \inst{1} \and
  J.P. Meunier   \inst{1} \and
  X. Regal       \inst{1} \and
  R. Sottile     \inst{1}
}

\institute{ 
Aix Marseille Universit\'e, CNRS, OHP (Observatoire de Haute Provence), Institut Pyth\'eas UMS 3470, 04870 Saint-Michel-l'Observatoire, France
\and 
  Laboratoire d’Astrophysique de Marseille, 
  Pôle de l’Étoile Site de Château-Gombert, 
  38, rue Frédéric Joliot-Curie, 13388 Marseille cedex 13, FRANCE \\
  \email{herve.lecoroller@lam.fr}
  \and 
  Universit\'e Nice-Sophia Antipolis, Observatoire de la C\^{o}te d'Azur,
  CNRS UMR 6525, BP 4229, F-06304 Nice Cedex, France
  \and 
  University of Namur (UNamur), research Center for Physics of Radiation and Matter (PMR-LARN), 61 rue de Bruxelles, B-5000 NAMUR
  \and 
  Lund Observatory, Box 43, SE-221 00 Lund, Sweden
  \and 
  Instituto de Astronom\'ia -- UNAM,
  Km 103 Carretera Tijuana Ensenada, 422860 Ensenada, B.~C., Mexico.
}

 \date{accepted October 04, 2014}


\abstract
   {The performance of interferometers has largely been 
     increased over the last ten years. But the number of observable
     objects is still limited due to the low sensitivity 
     and imaging capability of the current facilities.
Studies have been done to propose a new generation of interferometers.
}  
 %
   { The Carlina concept studied at the Haute-Provence Observatory consists in an optical interferometer configured as a diluted version of the Arecibo radio telescope: above the diluted primary mirror made of fixed co-spherical segments, a helium balloon or cables suspended between two mountains and/or pylons, carries a gondola containing the focal optics.
   This concept does not require delay lines.}
   {Since 2003, we have been building a technical demonstrator of this diluted telescope. The main goals of this project were to find the opto-mechanical solutions to stabilize the optics attached under cables at several tens of meters above the ground, and to characterize this diluted telescope under real conditions. In 2012, we have obtained metrology fringes, and co-spherized the primary mirrors within one micron accuracy. In 2013, we have tested the whole optical train: servo loop, metrology, and the focal gondola. 
 }
   {We obtained stellar fringes on Deneb in September 2013. In this paper, we present the characteristics of these observations: quality of the guiding,  $S/N$ reached, and possible improvements  for a future system.
}
   {By detecting fringes on Deneb, we confirm that the entire system
     conceptually has worked correctly. It also proves that when the
     primary mirrors are aligned using the metrology system, we can
     directly record fringes in the focal gondola, even in blind
     operation.
It is an important step that demonstrates the feasibility of building a diluted telescope using cables strained between cliffs or pylons. 
  Carlina, like the MMT or LBT, could be one of the first members of a new class of telescopes named Large Diluted Telescopes. Its optical architecture has many advantages for future projects: Planet Formation Imager, Post-ELTs, Interferometer in space.}


   \maketitle
%

\section{Introduction}

\begin{figure*}[!t]
  \centering
 \includegraphics[width=15cm]{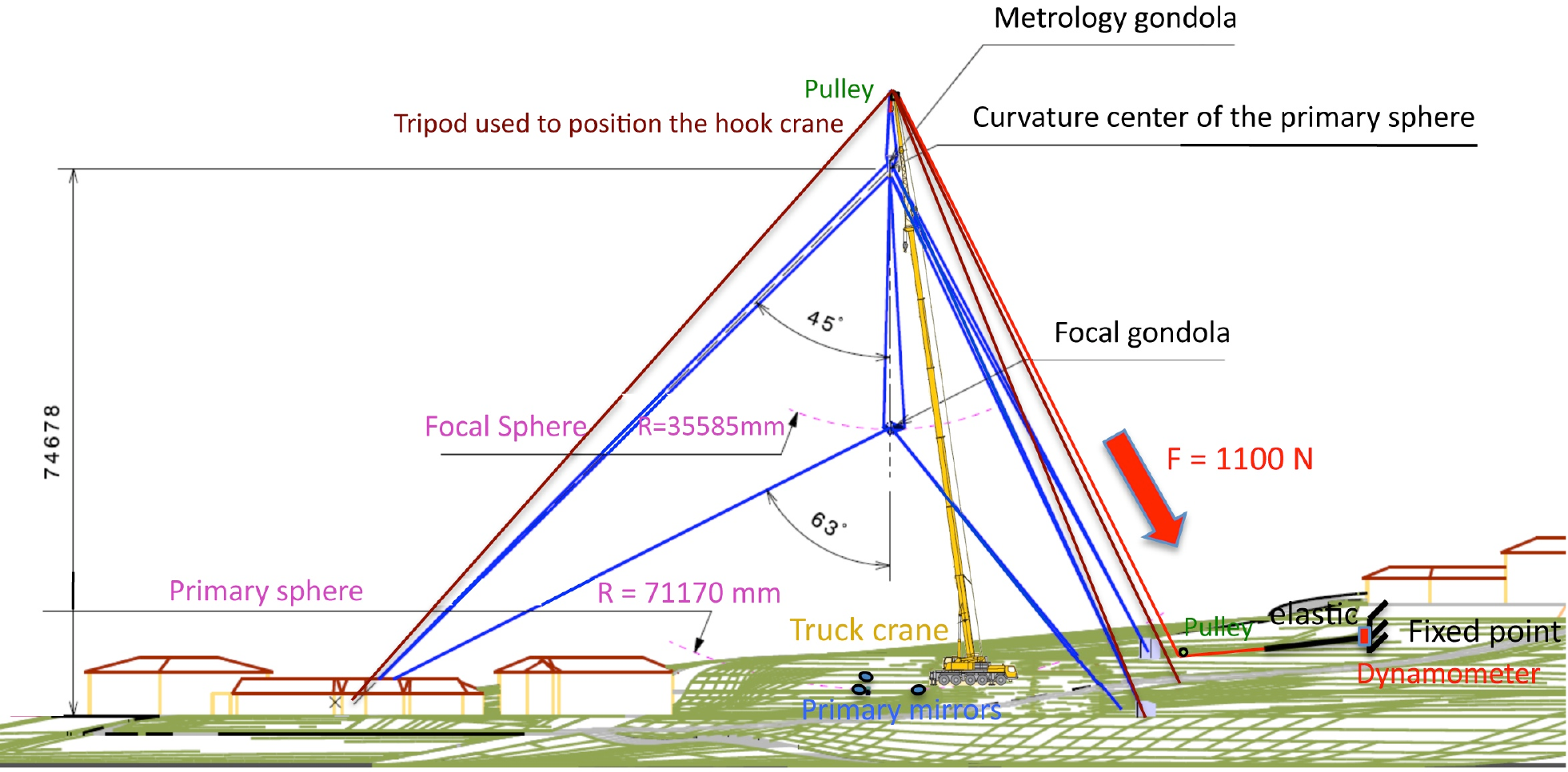}
  \caption{The Carlina-OHP experiment: In September 2013, the helium
    balloon (Paper I \& Paper II) was replaced by a truck crane. The
    thickness of the cables, and the diameter of the primary mirrors
    ($25\centi\meter$) have been exaggerated to be seen on this scale
    drawing.  An elastic rope attached to the ground passes through a
    pulley at the top of the crane and pulls on the same cable, with
    the same force (1100 N) as the helium balloon.}
 \label{crane}
\end{figure*}

The last ten years have seen a significant increase in scientific
publications making use of images based on optical
long-baseline interferometry. The sensitivity and imaging capability
of the interferometers have progressed. We are able to recombine up to 6 telescopes such as with the MIRC instrument on  CHARA \citepads{2010SPIE.7734E...2T}.
Nevertheless, the number of observable objects with this technique stays
limited due to the low sensitivity and the modest number
of telescopes we are able to recombine simultaneously (max. 6).
The current limitations of the interferometers lie also in their ability to detect highly contrasted features \citepads{2014ipco.conf..221M}.

Studies are under way to propose a new generation of interferometers, 
more  sensitive with higher imaging capabilities
than current systems. More than 6 telescopes are required
to reconstruct images of complex objects
\citepads{2012A&ARv..20...53B}.  They should be able to produce  images at very high contrast using coronagraphic techniques (e.g. \citeads{2014MNRAS.439.4018L}).
To reach this goal, the interferometers have to be improved along their full optical path. 
The number of reflections between the primary mirrors
and the focal instrument has to be minimized to limit the flux lost
in numerous
reflexions.
In this spirit, the OHANA team has studied the possibility to transport the light in optical fibers (\citeads{2000SPIE.4006..708P}; \citeads{2005AAS...207.8216W}).
The focal instrument can be optimized for example with integrated
optics with better throughput and with photometric calibration capability
\citepads{1999ASPC..194..264B}. It is crucial to damp vibrations of
sensitive parts, such as mirror mounts, delay lines, etc.\\
\noindent At the Haute-Provence observatory, we have studied a new concept for a diluted telescope. As described in two previous papers (\citeads[][Paper I]{2004A&A...426..721L}; \citeads[][Paper II]{2012A&A...539A..59L}), it is configured as a diluted version of the Arecibo radio telescope: above the diluted primary mirror made of fixed co-spherical segments, a helium balloon (or a crane) carries a gondola containing the focal optics. It works without delay lines and the focal gondola is the only moving part. With this simple optical train, in the future, tens or hundreds of mirrors could be added in the pupil to improve the imaging capability and sensitivity of the interferometer.\\
The biggest difficulty for this system is to stabilize the focal gondola that moves to track the stars, attached to cables at tens of meters above the ground. Indeed, the fringes move on the focal sphere with the Earth rotation and the focal optic has to track them exactly at the correct velocity to avoid that they drift on the camera during the exposure time. The speed and the maximum velocity drift of the focal gondola beyond which the fringes are totally blurred are of the same order of magnitude as the ones of a delay line of an interferometer having a baseline of equivalent length. This is a challenge for a cable-suspended focal gondola (see Table 1 of \citeads{2014ipco.conf..135D}). The main goals of this project were thus to find opto-mechanical solutions to stabilize the optics attached under cables, and to test this interferometer in real conditions, i.e. to obtain fringes that stay in the field of view of the focal optic, and that move slowly enough to be frozen in a short exposure ($\approx 5\,\milli\second$).
Note that the first fringes have been obtained in May 2004 with two closely-spaced primary segments providing $40\,\centi\meter$  baseline, and a CCD on the focal gondola (Paper I). These first encouraging results had proved that we can track the stars with a gondola attached by cables $35\,\meter$ above the ground. But, the two primary mirrors were easily co-spherized because they were adjacent. Moreover, the maximum velocity drift acceptable to freeze the fringes in a short exposure is proportional to the fringes size ($\lambda \frac{f}{B}$), i.e. to the aperture ratio $\frac{f}{B}=35/0.4=88$ in 2004.
A future scientific project would have much bigger baselines, and would work at a larger aperture ratio (typically $F/D=2-3$) to minimize the altitude of the focal gondola. In this case, the maximum velocity drift to freeze the fringes is 44 times smaller than in our experiment of 2004. In order to test the Carlina architecture in more realistic conditions than in 2004, we decided to validate the concept at $F/3.5$ (baselines of $5$--$10\,\meter$).\\

\begin{figure*}[!t]
  \centering
  \includegraphics[width=17cm]{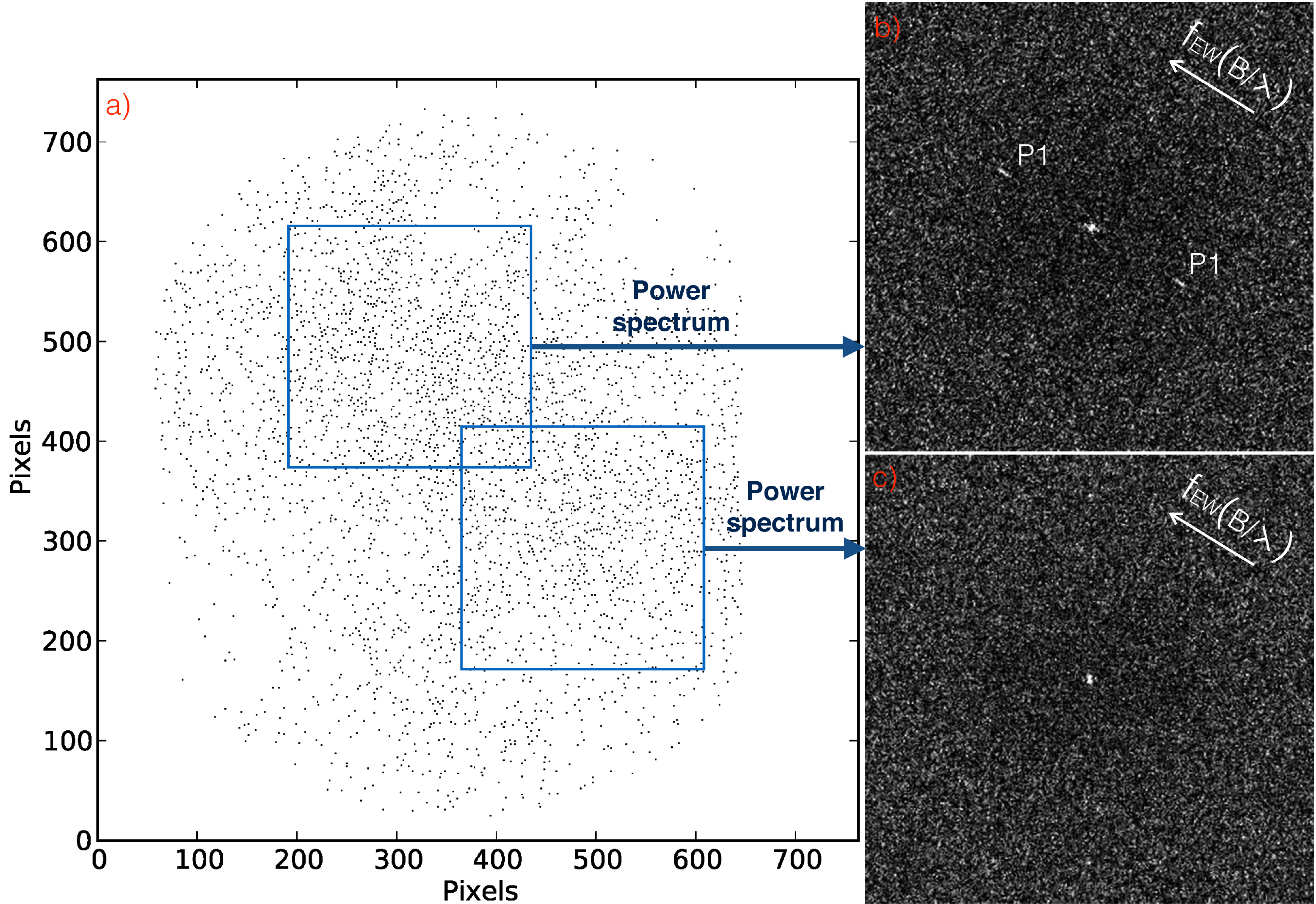}
  \caption{a) Binary image (cropped to $761\times761$ pixels) obtained on Deneb in $1\, \milli\second$ exposure time with the Carlina photon counting camera. The light that falls on the CCD comes from the intensifiers of circular shape. This image contains 3524 Photons. b) Square modulus of the Fourier Transform of the top left image (blue square of $245 \times 245$ pixels). In this part of the image, we detect the fringe peaks of the $5\, \meter$ baseline (P1 mark). The intensities in the power spectrum are represented using a linear scale. The white arrow indicates the direction and the norm of the spatial frequency of the $5 \meter$ baseline (oriented East-West) at $510\, \nano \meter$. Only two images among the Deneb data allow a direct detection of the fringes (in an individual frame without adding several power spectra).  c) No fringes are detected in the power spectrum of the bottom right image.  i.e. the fringes are located in the top left part of the image a).}
  \label{image_photons}
\end{figure*}

\noindent In 2011-2012, we obtained metrology fringes, and we co-spherized distant primary mirrors (baselines $5$--$10\,\meter$) within one micron accuracy (Paper II). The opto-mechanical design, the metrology system and the servo loops have been largely described in Paper I, Paper II, and in several conference proceedings (\citeads{2012SPIE.8445E..14L}; \citeads{2014ipco.conf..135D}). \\
\noindent In this paper, we  focus on the observation performed in September 2013. In Sect. \ref{characteristic}, we recall quickly the main prototype characteristics that are relevant for the data reduction (baseline, photon counting camera, guiding camera characteristics, etc.). In Sect. \ref{observations}, we present the data recorded in September 2013. Data reduction, and fringes on Deneb are shown at Sect. \ref{reduction}. The guiding behaviour is described in Sect. \ref{guidage}. A discussion is provided in Sect. \ref{discussion} and the conclusion is given in Sect. \ref{conclusion}.


\section{Carlina prototype characteristics} \label{characteristic}
\label{characteristic}

\begin{figure*}[!t]
  \centering
  \includegraphics[width=18cm]{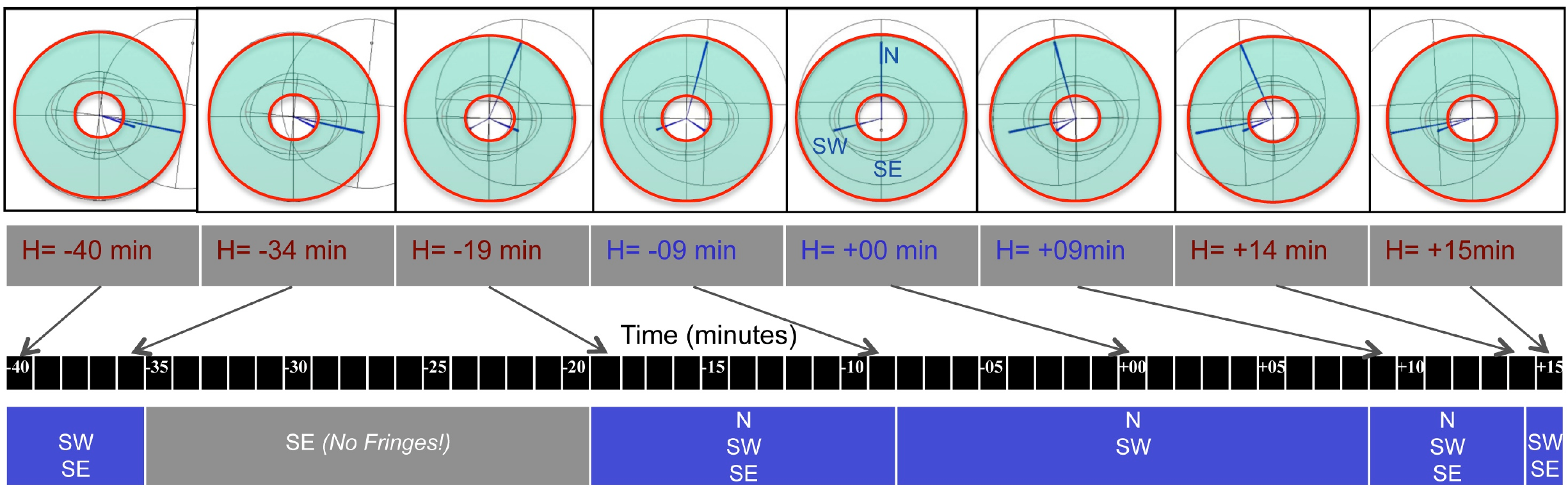}
  \caption{Top: Zemax ray-tracing software views from above the focal
    gondola on the optical axis. These graphs show the optical rays
    (blue lines) for the declination of the star Deneb at several hour
    angles ($-40\,\minute <H<+15\,\minute$).  In this projected view, each
    mirror (N, SW, SE) is at the end of a blue line. It falls in the
    Mertz field of view if it is located between the two red circles
    (in the blue area). The central obscuration is due to the central hole of the Mertz corrector (see Fig.~2 \&
    Fig.~4 of Paper I): when a primary segment is aligned with this
    hole (i.e. the focal gondola is aligned with this segment and the star), the light is lost (the lost light can be blocked
    by a mask in the pupil plane, to avoid directly illuminating the camera).  A
    graph is shown each time a primary mirror enters in or exits from the Mertz field of view. Note that the North mirror is represented, but during the
    observation of Deneb it was not aligned, and was masked. Thus, we expected fringes on Deneb only when the SW and SE mirrors are in the field of view.
      Bottom: We give the name of the primary mirrors (N,
    SW, SE) that are in the field of view of the Mertz corrector
    during the tracking of Deneb. From $35\,\minute$ to $20\,\minute$ before
    the transit only the SE mirror is in the field of view of the
    Mertz corrector, and no fringe acquisition is possible. We have
    recorded data on Deneb during $1\,\minute \, 30\,\second$ starting $10\,\minute$
    before the transit (transit at $H=00\,\minute$
    i.e. 21:18:28 UT on September 05, 2013), and during
    $5\,\minute$, $9\,\minute\, 30\,\second$ after the transit (see the observing
    log of Table \ref{tabllog}). During $18\,\minute$ around the transit,
    the SE mirror of the small baseline is not in the field of view of
    the Mertz corrector (confirmed during the observation: Table
    \ref{tabllog}).}
  \label{fview}
\end{figure*}

Carlina can be divided into four blocks: The primary mirrors, the focal gondola, the metrology gondola and the holding system (Fig. 1 of Paper II). Carlina specifications are provided in Table A.1 of Paper II, and in Table 1 of \citetads{2014ipco.conf..135D}. Here, we just recall the main characteristics of the prototype built at the Haute-Provence observatory: \\
\noindent  The primary mirror is made of three segments positioned on a 71.2 meter radius sphere (Fig. \ref{crane}), forming three baselines of respectively 5, 9 and 10.5 meters (Paper II). The metrology gondola (Paper II), and the focal gondola are then respectively at $71.2\,\meter$ and $35.6\,\meter$ above the ground. From 2003 to 2012, the metrology and focal gondolas were carried by cables attached under a helium balloon. But experience proved that the use of a helium balloon exhibits a serious drawback: it requires a complete disassembly of the installation every morning, and a re-assembly every evening! The entire instrumental setup using the balloon was not designed to resist to winds above $20\,\kilo\meter\per\hour$. The balloon has thus been replaced by a lifting crane (Fig. \ref{crane}) for the last observation run, in September 2013 \citepads{2014ipco.conf..135D}. This holding system behaves as the pylons that could be used for a future project (Paper II).\\

\noindent In 2013, we have also simplified the focal gondola (\citeads{2014ipco.conf..135D}) by removing the pupil densifier \citepads{1996A&AS..118..517L}. The required guidance accuracy ($0.1\, \rm arcsec$) was indeed impossible to reach without an embedded fine guiding system. The Fizeau focal recombiner used to record the data presented in this paper is only equipped with a corrector of spherical  aberration, and guiding and science cameras. A dichroic mirror reflects the red part of the light onto the guiding camera ($\lambda > 700\,\nano\meter$) and lets the blue light ($\lambda <700\,\nano\meter$) pass toward a photon-counting science camera \citepads{Blazit2008}. The diameter of the field of view of the science camera  is 2 arcsec. The field of view of the guiding camera (AVT Guppy F-038 NIR from Allied Vision Technologies company) is 40 arcsec. The science camera uses two intensifiers \citep{Blazit2008}. The intensifiers feed a $1024\times 1024$ pixels CCD \footnote{MV-D1024-CL CMOS model from the Photon Focus company}. The guiding and science  images are sent to a ground-based  PC by optical fibers with a maximum transfer rate of 33 science images per second. The science images are cropped to 761x761 pixels corresponding approximately to the size of the circular intensifiers (Fig. \ref{image_photons}).

\section{Observing log}\label{observations}

We have observed three stars during a run of 4 nights from September 2 to 5, 2013: $\gamma\, And$, Deneb ($\alpha\, Cyg$), and $o\, And$.
Fringes were obtained on Deneb during the night of the run with the best atmospheric conditions (Table \ref{tabllog} ).
A green filter centred at $\lambda =510\, \nano\meter$ with $84 \, \nano\meter$ bandwidth has been used at the entrance of the science camera.
 The exposure time of the science camera was adjusted between $1\,\milli\second$ and $10\,\milli\second$. 
With only three mirrors (Paper II, and \citeads{2014ipco.conf..135D}) our prototype can observe only near Zenith (declination: $40 \deg<\delta<50 \deg$). 
Moreover, the duration of observation depends on the declination of the star. Fig. \ref{fview} shows the observing windows for Deneb.\\
 \noindent We also recorded images with the guiding camera to evaluate the stability of the focal gondola during tracking.
 

\section{Data reduction} \label{reduction}

The binary images (Fig.~\ref{image_photons}) of the science camera were transformed into FITS
format using a C++ script calling routines from the
\emph{quarklib} library \footnote{The library is currently being
  reimplemented as a C++ module for \texttt{node.js}, see
  \url{https://github.com/Nunkiii/node-fits}}.

\begin{figure*}[!t]
  \centering
  \includegraphics[width=16cm]{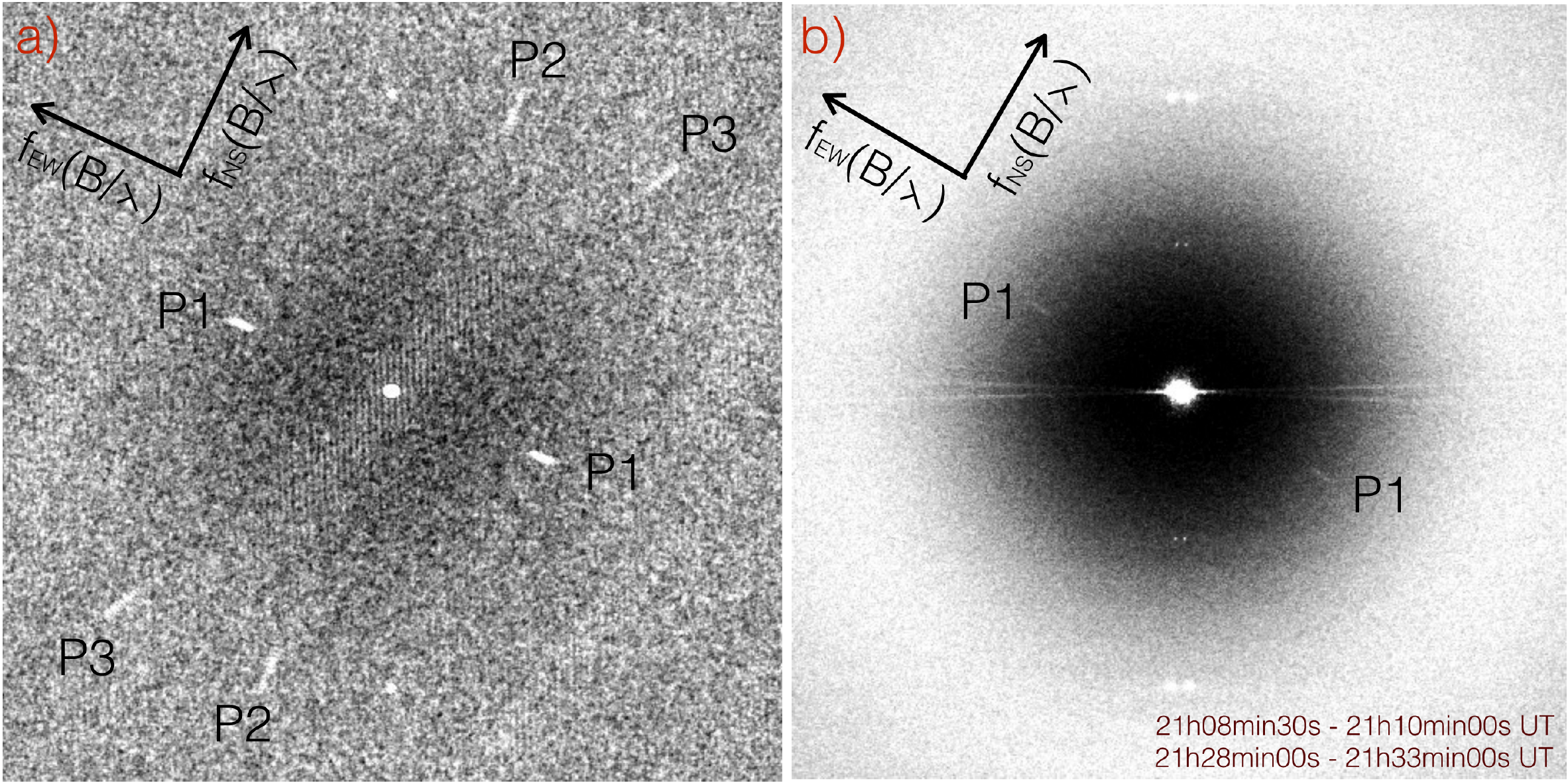}
  \caption{a) Power spectrum of the Fizeau laboratory fringes obtained using a supercontinuum laser source in the focal plane of the Mertz corrector 
  and a mask positioned on a lens just after (see L1 Lens in Fig. 3 of \citeads{2012SPIE.8445E..14L} and Fig. 12 of \citeads{2014ipco.conf..135D}).  
  The diameters of the three holes and their positions in the mask are adjusted using the Zemax ray tracing code to simulate
   the three stellar beams coming from the primary mirrors. The mask has been oriented to
    simulate the stellar fringes at the transit. P1, P2 and P3
    indicate the position of the fringe peaks, respectively for the
    $5\,\meter$, $9\,\meter$ and $10.5\,\meter$ baselines. The
    vertically aligned dots are an artefact of the science camera
    (visible even if the camera is illuminated uniformly without
    fringes).  b) Sum of all the power spectra computed for Deneb (3 000 images  recorded before the transit and 10 000
    images recorded after the transit: see log of the observations
    in Table \ref{tabllog}). The Earth
    rotation correction has turned the "artefacts" of the images taken before the transit 
    counterclockwise, while they turned clockwise for the images taken after the transit. The intensities are represented using a linear scale.
    We see the Deneb fringe peaks of the $5\,\meter$
    baseline (P1). They are at the expected position with the same shape
    as in the laboratory (image a) using the green filter
    ($\lambda=510\,\nano\meter$, $\Delta\lambda = 84\,\nano\meter$).
 The black arrows define the coordinate system of the panel. The $f_{EW}$ arrow indicates the direction and the norm of the spatial frequency of the $5 \meter$ baseline (oriented East-West) at $510\, \nano \meter$ while the $f_{NS}$ arrow shows the direction of the spatial frequency of the $10 \meter$ baseline (oriented North-South). It has the same size as the $f_{EW}$ arrow.
}
  \label{deneb}
\end{figure*}

\noindent During $1\, \hour$ around the transit, the motion  of the fringe peak can be approximated by a counterclockwise linear
rotation of $\approx 0.68\,\degree$ per degree of hour angle ($\approx 0.17\,\degree$ per min). 
Moreover, the width of the peak is proportional to the diameter of the primary mirrors (\citeads{1984JOpt...15..363R}, \citeads{1989dli..conf..125R}). It has been estimated in the image by using the power spectrum simulated in the laboratory (Fig. \ref{deneb}) giving 
 $\approx 3 \,\degree$ view from the center of the power spectrum i.e. it turns $1/10$ eme of its width in about $ 2\, \minute$.
So, we compute the power spectrum of a series of images by taking
the sum of the squared modulus of the Fourier transform of 1000 frames
($\approx 30$ seconds of acquisition) without rotation correction.  
Over a larger period, we align the fringe peaks (each one computed with 1000 images without rotation correction) by applying a rotation (without
interpolation). Finally, we co-add all the computed power spectra.

\noindent A fringe peak is only detected on Deneb (Fig. \ref{image_photons} \& Fig.~\ref{deneb}) on September 5 between 21:08:30 UT and 21:33:00 UT on the $5\,\meter$
baseline (that night, we worked only with two
mirrors; the North aperture was closed). 
The fringe peak has a $S/N=5.4$ (Table \ref{SN}) and it is at the expected shape and
position. It is slightly elongated due to the relatively large
bandwidth filter ($\Delta \lambda=84\,\nano\meter$) that we used (see Fig.~\ref{deneb}).

\subsection{Signal to noise ratio}

We estimate the signal to noise ratio of the fringe peak as:

\begin{equation}
  \frac{S}{N}=\frac{I_{Peak}-<I_{Bg}>}{\sigma_{Bg}}
\end{equation}
where:
\begin{itemize}
\item $I_{Peak}$ is the intensity integrated in the fringe peak;
\item $<I_{Bg}>$ is the average intensity of the background around the fringe peak.
\item $\sigma_{Bg}$ is the standard deviation of the background around the fringe peak.
\end{itemize}
Table \ref{SN} gives the $S/N$ computed for the fringe peaks presented
in Fig.~\ref{deneb} b). The highest signal to noise ratio ($S/N=6.6$) of the fringe peak is obtained 
if we remove the data from 21:30:00  to 21:33:00, showing that there is no signal during this period. \\

\begin{table*}[!t]
\centering
 \caption{Signal to noise ratio of the fringe peak for Deneb}
    \begin{tabular}{ccc}
\hline
\hline
        Observing periods & Number of  images &$S/N$  \\
               (hh:mm:ss) & &(with rotation correction)\\
\hline
            21:08:30--21:10:00&3000& 4.7    \\
            21:28:00--21:33:00&10 000 & 4.13 \\
21:08:30--21:10:00\, +\, 21:28:00--21:33:00 &13 000 &5.4 \\
21:08:30--21:10:00\, +\, 21:28:00--21:30:00 &7000& 6.6 \\
\hline
\end{tabular}
    \label{SN}
\end{table*} 

\subsection{Visibility} \label{visibility}

\begin{figure}[!t]
  \centering
  \includegraphics[width=8.8cm]{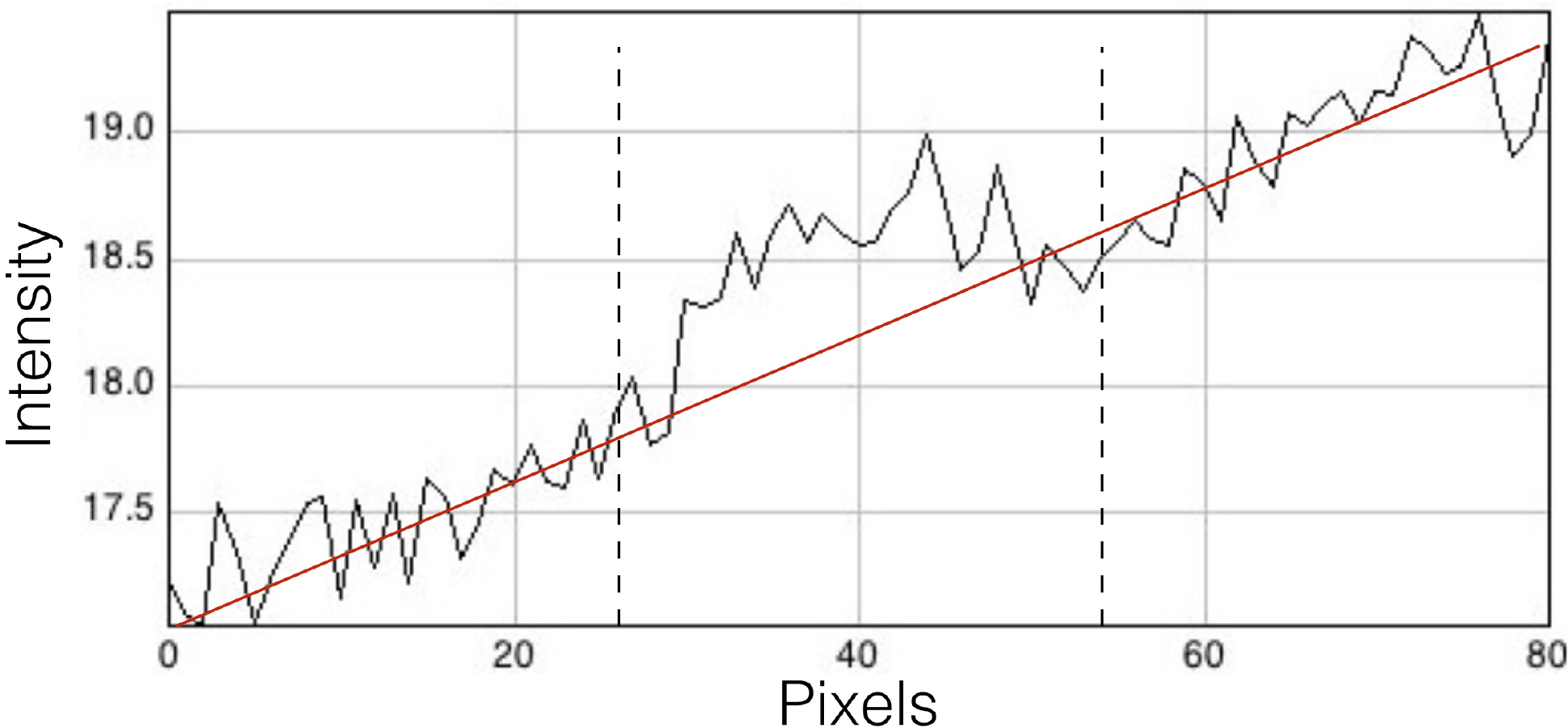}
  \caption{Section view along the fringe peak of Deneb (P1 in the $f_{EW}$ direction of Fig. \ref{deneb}) using data S/N=6.6 (Table \ref{SN}). To show the energy over the entire peak, a binning on its width (3 pixels in the FFT sampling) has been done before plotting. To extract a visibility, the energy of the peak is integrated between the two vertical dashed lines and above a polynomial fit.}
  \label{coupe_frange_Deneb}
\end{figure}

\begin{figure}[!h]
  \centering
  \includegraphics[width=8.8cm]{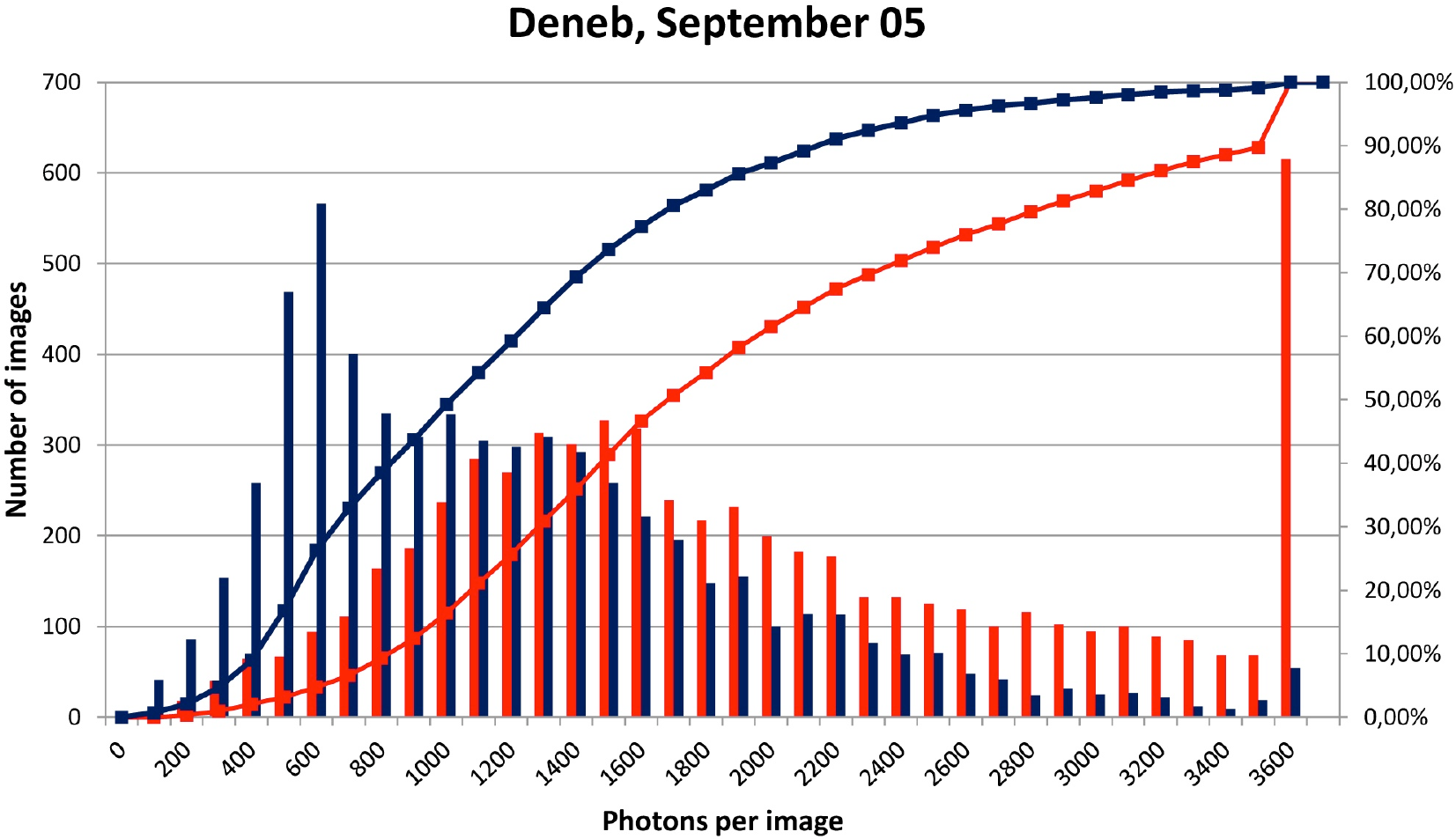}
  \caption{Histograms of the photons per image.  The line curves show the cumulative percentage of
    images of the histogram of the same color. No fringes were detected for the data of the blue histogram
    (data between 21:30:00 - 21:33:00 UT of
    Deneb). The red histogram correspond to the  data where the fringes have
    been detected (21:08:30-21:10:00 UT +
    21:28:00 - 21:30:00 UT).  The large number of images at 3600 photons in the red histogram shows that Deneb slightly saturates  the science camera. 
   For the data of Deneb with
    fringes (in red), $45 \%$ of the images have more than 1800  photons
    (1800 is a little below half of the maximum flux on Deneb that is over the saturation level),
    while only $25 \%$ of the data without fringes (blue) exceed this
    value.}
\label{histo}
\end{figure}

We have extracted a visibility from the power spectrum of
Deneb  using data recorded between
21:08:30 and 21:30:00 UT: $V_{Deneb}(5\,\meter)=
0.03$. The visibility measured on the single image shown at Fig. \ref{image_photons} gives $V_{Deneb}(5\,\meter)=0.3$. These values are much smaller than the expected value of
$V_{UD}(5\,\meter)= 0.99$. But, the visibility measurements are 
strongly degraded by atmospheric turbulence, and should be
calibrated on a reference star. This is
particularly relevant with our simple recombiner that works without spatial
filtering, tip-tilt and photometric correction. The  saturation
of Deneb (Fig.~\ref{histo}) may also significantly decrease the visibility by removing
more photons from the bright than the dark fringes 
\citep{Blazit2008}. Nevertheless, the goal here is simply
to show that we are able to integrate the flux in the fringe peak
using the \citetads{1998A&AS..129..609B} method (to remove the so-called photon-centroiding hole) to measure a
visibility point (Fig.~\ref{coupe_frange_Deneb}).  However, much more data (than 7000  images of $1\,\milli\second$
exposure time i.e. $7\,\second$ of integrated signal)  would be required to compute a mean visibility
with a good signal to noise ratio.

\begin{figure}[!h]
  \centering
   \includegraphics[width=8.8cm]{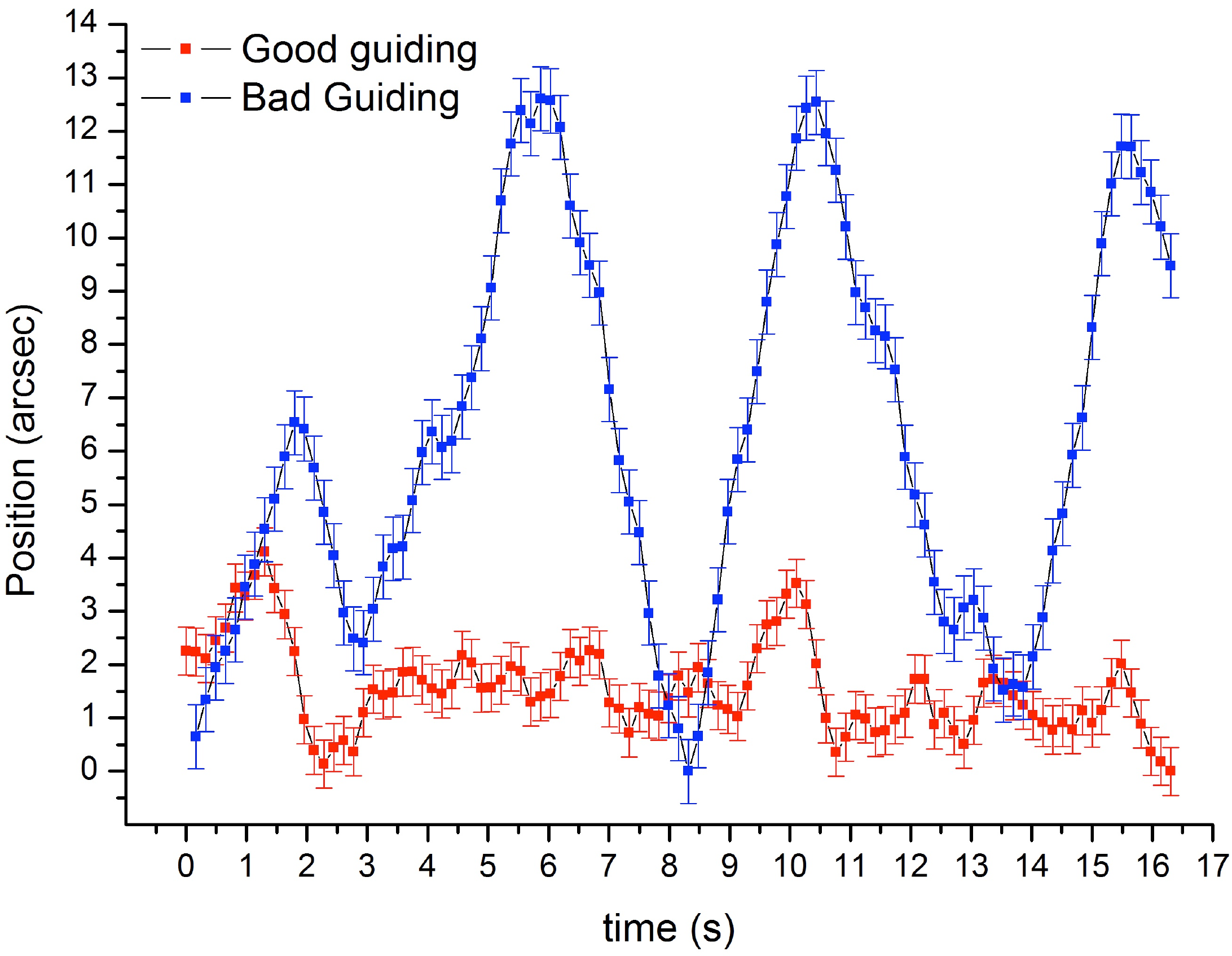}
  \includegraphics[width=8.8cm]{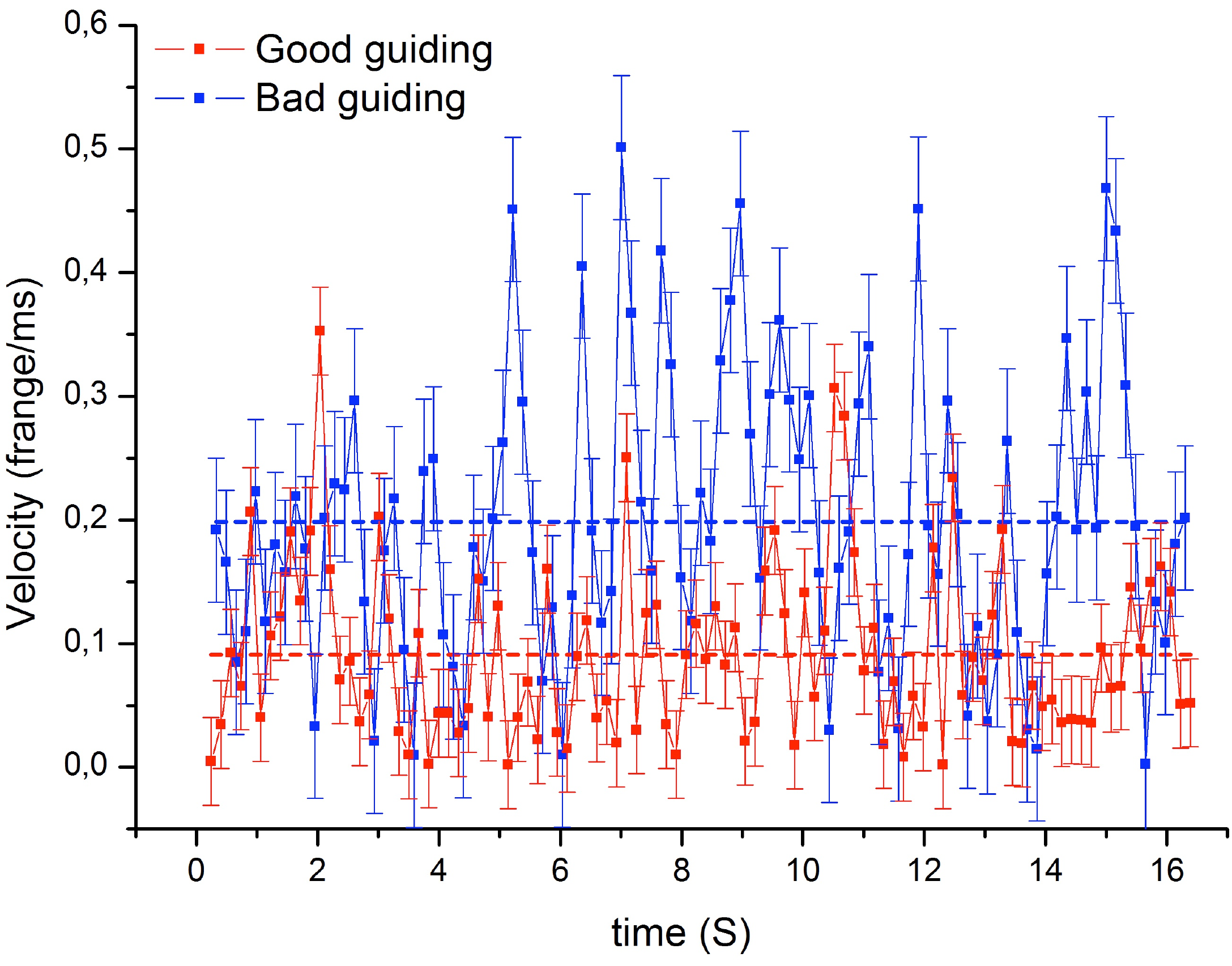}
  \caption{Position (top), and its derivative (velocity at bottom) of
    the guiding focal image projected on an axis perpendicular to the fringes of the small baseline.
      The red and blue curves
    correspond to periods with small and large amplitude oscillations obtained on respectively
    $o\, And$ (data selected among the recorded images between 22h57
    and 23h00) and $\gamma\, And$ (data selected in the first minute
    of the recorded images between 02h52 and 02h57).   The unit of the vertical position axis is one
    arcsecond (1 pixel $= 0.133''$ on the guiding camera).  The
    error bars are
    the median "seeing" at the moment of the observation. The error bars of the bottom graph are the
    standard deviation ($\sigma$) of the velocity points.  The mean velocities are of 0.2
    and 0.1 fringe$/\milli\second$  for  
    guiding with large and small oscillations, respectively. One pixel of the guiding camera equals 6 fringes for the
    $5\,\meter$ baseline. It is expressed in fringe$/\milli\second$ but
    for guiding with small amplitude oscillations, the fringes move probably less than 0.1 fringe$/\milli\second$ because this value is
    affected by the atmospheric tip-tilt.}
\label{Positionvelocity}
\end{figure}

\section{Guiding characterisation} \label{guidage}

\subsection{Analysis based on the science camera images}

\noindent  We have recorded
data for Deneb continuously  from
21:08:30 to 21:10:00 and from 21:28:00 to 21:33:00 but we do not detect fringes from 21:30:00 to 21:33:00 (6000 images).
To analyse why the fringe peak is not
present during all of the observations,  we use the following property: the field of view of the science camera ($\approx
2\,\arcsecond$) has the same size as the typical seeing at
OHP. The flux thus decreases immediately when the image becomes off-axis. \\

\noindent Fig.~\ref{histo} shows in red the histogram
of the data where we detect the fringes (21:08:30--21:10:00 UT +
21:28:00--21:30:00 UT; 7000 images), while the blue histogram is computed with
the data without fringes signal (21:30:00--21:33:00 UT). Clearly,
the red histogram shows more images at high flux than the blue
one. $45 \%$ of the images of the data with fringes detected have
more than 1800 photons while only $25 \%$ of the data without fringes
 exceed this value. Note that the histograms obtained on $\gamma \, And$ and $o\, And$ have approximately the same shapes as the blue histogram on Deneb:
 $85-95\%$ of the recorded images have less than $50 \%$ of their maximum
flux, i.e.  most of the time the
fringes were not detected because the light was out of the science
camera due to bad guiding.  But it is not the only reason. In the next section, we show that the fringes are also blurred by guiding oscillations.

\subsection{Analysis based on the guiding camera images}

Guiding has been evaluated with only one aperture feeding the recombiner (Table \ref{tabllog}).\\

\noindent The top plot of Fig.~\ref{Positionvelocity} shows, during 
small and high amplitude guiding oscillations
periods, the position of the focal image projected
on an axis perpendicular to the fringes of the $5\,\meter$ baseline, i.e. along
the guiding CCD width, considering the optical design of the focal
optics.  Fig.~\ref{Positionvelocity} shows that the projected motion
(perpendicular to the fringes) for a guiding period with large amplitude oscillations can be well
approximated by straight lines with abrupt (practically instantaneous)
changes in direction.  Also, the velocity mean is linked to this
global motion, and it is weakly affected by the atmospheric tip-tilt.

The mean velocity when guiding with large amplitude oscillations (blue curve in
Fig.~\ref{Positionvelocity}) is 0.2 fringe/ms (1 fringe with an exposure
time of 5 ms).  Fringes obtained with longer 
exposure times ($\geqslant 5\,\milli\second$) are totally blurred. Note that during our observations, most of the time we were guiding with large amplitude oscillations ($\approx 12\, \arcsecond$) and we
adjusted the exposure time to 5-10 ms (Table \ref{tabllog}), explaining why we did not detect fringes on $\gamma\, And$ and $o\, And$. 
Moreover, each night $\gamma \, And$ and $o\, And$ have been observed with average wind speeds of
$0.25-2.6 \,\meter\per\second$ (Table \ref{tabllog}).  Deneb is the only star that
has been observed during a long period (20h30 - 21h45 UT) without any
wind and using very short exposure time ($1\,\milli\second$). The best seeing condition
($0.9\,\arcsecond$) of the observing run was also obtained during that
observation of Deneb.


\section{Discussion} \label{discussion}

We have shown above that the low $S/N$ of the detected fringe peak can largely be explained by guiding errors. 
Our prototype indeed works without tip-tilt correction, and we
     track the stars by winch control on $100\,\meter$ long cables. 
     Nevertheless, the measured guiding oscillations ($\approx 12\, arcsec$ at
0.2 Hz) can be easily corrected by adding tip-tilt correction. The $S/N$
could be also improved by using a photon-counting
camera with a much smaller readout time. The effective integration was only $1/30$ of the observing time (duty cycle 0.03), i.e. the $S/N=6.6$ (Table \ref{SN}) corresponds to $7\,\second$ of integrated
time.

\noindent  Our results confirm that the stability of the gondola suspension and the associated control system are critical. 
More studies (theoretical and experimental) are required to better characterize high frequency vibrations, in particular if we wish to use this interferometer for astrometry, or with an adaptive optics system (adapted to a diluted telescope) for coronagraphy. Cable models have been set up (\citeads{2011SPIE.8336E..18E}, \citeads{2014ipco.conf..153A}).  A theoretical study based upon an exact numerical model of the Carlina experiment and including all cables and suspended masses is in progress and will be documented separately \citep{Andersen2014}.

  
\section{Conclusions}\label{conclusion}

   \begin{enumerate}
   \item Fringes have been obtained on Deneb with $5\,\meter$ baseline
     using a filter of $84\,\nano\meter$ bandwidth, centred at $\lambda
     = 510 \,\nano\meter$.
   \item The results confirm that the entire system conceptually has worked
     correctly. They also show that, when the primary mirrors are
     aligned using the metrology system (Paper II), we can directly
     record fringes in the focal gondola, even in blind operation.
     \noindent The low signal to noise ratio ($S/N=6.6 $) can largely
     be explained by oscillations of the focal gondola. 
     Modeling suggests that if the cable preload is too low, then a
     gain in suspension stiffness of an order of magnitude should be
     possible by increasing the preload. 
      It is also
     believed that a much better gondola stability could be obtained
     using servo-controlled reaction wheels, and reaction masses on
     board the gondola with appropriate angular and linear gondola
     acceleration, velocity and position feedback. Hence, considering
     these potential future improvements, we find the above results highly encouraging.
   \item The Carlina-type diluted telescope concept could be
     interesting for an interferometer with high imaging capability
     and sensitivity (for example: $>10$ mirrors of $1$--$5\,\meter$ each;
     baseline $50$--$100\,\meter$; near IR range; pylons to carry the optics).
     It could perhaps be installed at an astronomical site by digging an artificial
     crater. More studies are required to determine the
     cost of such a solution.  It could be combined with auxiliary
     telescopes at very long baselines (for example: $0.5$--$ 1\,\kilo\meter$) by
     using delay lines. Such an uv coverage (dense on the shorter
     baselines, and more diluted on the longer ones) is similar to the
     ALMA radio telescope geometry, and is optimal for imaging of
     complex objects. This solution
     could be studied for the Planet Formation Imager project\footnote{
  \url{http://planetformationimager.org}} proposed by \citet{Monnier2014} during the OHP2013 colloquium
     (\citeads{2014ipco.conf..277S}).
     Such an interferometer could also be attractive for post-ELTs
     $50$--$100\,\meter$ diluted aperture. In space, the
     metrology from the center of curvature (Paper II) could be used to
     cophase the spherical primary mirrors with an extremely high
     accuracy.
   \end{enumerate}

\begin{acknowledgements}
  This research has been funded by ASHRA, CNRS/INSU, Pytheas/OSU and
  Coll\`ege de France. Mechanical elements were built by the technical
  group at OHP and Nice observatory. Thanks to Mette Owner-Petersen for having analysed the 
  theoretical prospects of using  passive shock absorbers.
   Thanks to Jean Surdej, Stephane
  Dumont, and Armand Rotereau for their wonderful pictures of the
  experiment. We are grateful to the students and to the people who helped
  us during the long nights of tests: Julien Chombart, Jean-Philippe
  Orts, Romain Pascal, Sandrine Perruchot, etc. We are very grateful
  to the Rouvier Lafont Manosque company that sponsored the last night
  (loan of the truck crane) when we obtained Fringes ! We are
  grateful to the LEUKOS company that helped us to use their
  supercontinuum laser.
\end{acknowledgements}


\bibliography{LeCoroller}   
\bibliographystyle{aa} 

\onecolumn
\begin{appendix} 

 \section{Logbook of the observations}  \label{log}
 
The weather conditions reported in the last two columns of Table \ref{tabllog} have
been recorded \footnote{Seeing is measured with a SBIG seeing Monitor, see
  \url{https://www.sbig.com/products/cameras/specialty/seeing-monitor/}} by the OHP meteo station located 300
m north of the experiment. Note that the anemometer is not sensitive to wind below $1\, \meter \per \second$.
The values smaller than $1\, \meter \per \second$ are the  average wind speeds over the observation period given in the UT column.
Moreover, the anemometer is located close to ground, and it is likely that the wind was
somewhat stronger (due to a wind gradient) at the level of the Carlina
cables. Fringes on Deneb on Sept 5. were obtained during the most quiet period of the full observing run.

\begin{table*}[!h]
 \caption{Log book of September 2013 observations}
  \centering
  \footnotesize
\begin{tabularx}{16cm}{|p{0.8cm}|p{0.8cm}|p{1.3cm}|p{1.8cm}|p{1.3cm}|p{0.5cm}|p{0.7cm}|p{1.5cm}|p{1cm}|X|}
\hline
\textbf{Star name}  &\textbf{mV}&\textbf{Images number}&\textbf{Date} \newline (month day)&\textbf{UT} \newline (hh:mm:ss)&\textbf{Exp. time\newline  ($\milli\second$)}&\textbf{Bases}\newline ($\meter$)&\textbf{Camera type}&\textbf{Av. \newline seeing}  \newline (arcsec.)&\textbf{Av. \newline Wind speed} \newline ($\meter\per\second$) \\
\hline
      Lab. source       &       ...       & 10000 &     July 19     &         ...           & $15$  & 5 \newline 9 \newline 10.5 & science  (1) &... &...\\
\hline 
    $\gamma\, And$      &      2.10     &  40000 &    Sept. 02     & 02:47:00  03:24:00 & $10$ & 5 \newline 9 \newline 10.5 & science &1.21 &0.48  \\
\hline   
    $\gamma\, And$      &      2.10     & 70000 &     Sept. 03     & 02:26:00  03:09:00 &  $5$  & 5 \newline 9 \newline 10.5 & science  (2) &1.2&0.25  \\
\hline   
    $\gamma\, And$      &      2.10     &  40000 &     Sept. 04     & 02:28:00  02:50:00 &  $5$  & 5 \newline 9 \newline 10.5 & science &1.27&2.6 \\
\hline 
 Deneb ($\alpha \, Cyg$) &      1.25     & 3000   &     Sept. 05     & 21:08:30  21:10:00 & $1$  & 5 & science &0.91&0 \\
\hline 
 Deneb ($\alpha \, Cyg$) &      1.25     & 10000 &     Sept. 05     & 21:28:00  21:33:00 & $1$  & 5 & science &0.89&0 \\
\hline    
    Sky background      &       ...      &  5000 &     Sept. 05     & 22:45:00 22:47:30 & $ 5$  & 5 & science &...&... \\
\hline    
      $o \, And$        &    3.62     &1113&     Sept. 05     & 22:57:00  23:00:00 &  $100$    & - & guiding  &...&...  \\
\hline     
     $o \, And$        &      3.62     & 10000 &     Sept. 05     & 23:15:30  23:20:30 &  $5$  & 5 & science &1.00&1.1  \\
\hline      
      $o \, And$        &      3.62     & 1253  &     Sept. 05     & 23:44:00  23:48:00 &  $100$    & - & guiding &...&...  \\
\hline    
    $\gamma\, And$      &     2.10     & 30000&     Sept. 05     & 02:19:30  02:35:30 &  $5$  & 5 & science &1.21&0.66  \\
\hline    
    $\gamma\, And$      &     2.10     &  1524  &     Sept. 05     & 02:52:00 02:57:00 &   $100$    & - & guiding &...&... \\
\hline
\end{tabularx}
      \tablefoot{The first line
        corresponds to a laboratory observation done in July 2013. The
        dates correspond to the beginning of the night. The UT column
        gives the starting and ending time of recording.  
        \small \emph{Remarks:
        (1) Laboratory Fringes; 
        (2) Only NS base ($9\,\meter$) at the end.}}
      \label{tabllog}
  \end{table*} 
  
\end{appendix}
\end{document}